\documentclass[twocolumn,reprint,aip,jap,amssymb, superscriptaddress]{revtex4-1}
\usepackage{amsmath}
\usepackage{amssymb}
\usepackage{amsthm}
\usepackage{graphicx}
\usepackage[english]{babel}  
\usepackage{color}
\usepackage{enumerate}

\begin{document}

\title{Observation of nonlinear bands in near-field scanning optical microscopy of a photonic-crystal waveguide}

\author{A. Singh} 
\affiliation{Complex Photonic Systems (COPS), MESA+ Institute for
Nanotechnology, University of Twente, PO Box 217, 7500 AE Enschede, The Netherlands}
\affiliation{Optical Sciences (OS), MESA+ Institute for
Nanotechnology, University of Twente, PO Box 217, 7500 AE Enschede, The Netherlands} 

\author{G. Ctistis} \email{g.ctistis@utwente.nl}
\affiliation{Complex Photonic Systems (COPS), MESA+ Institute for
Nanotechnology, University of Twente, PO Box 217, 7500 AE Enschede, The Netherlands}

\author{S.R. Huisman}  
\affiliation{Complex Photonic Systems (COPS), MESA+ Institute for
Nanotechnology, University of Twente, PO Box 217, 7500 AE Enschede, The Netherlands}
\affiliation{Optical Sciences (OS), MESA+ Institute for
Nanotechnology, University of Twente, PO Box 217, 7500 AE Enschede, The Netherlands} 

\author{J.P. Korterik}
\affiliation{Optical Sciences (OS), MESA+ Institute for
Nanotechnology, University of Twente, PO Box 217, 7500 AE Enschede, The Netherlands} 

\author{A.P. Mosk}
\affiliation{Complex Photonic Systems (COPS), MESA+ Institute for
Nanotechnology, University of Twente, PO Box 217, 7500 AE Enschede, The Netherlands} 

\author{J.L. Herek}
\affiliation{Optical Sciences (OS), MESA+ Institute for
Nanotechnology, University of Twente, PO Box 217, 7500 AE Enschede, The Netherlands} 

\author{P.W.H. Pinkse}  \email{P.W.H.Pinkse@utwente.nl}
\affiliation{Complex Photonic Systems (COPS), MESA+ Institute for
Nanotechnology, University of Twente, PO Box 217, 7500 AE Enschede, The Netherlands}


\begin{abstract}
We have measured the photonic bandstructure of GaAs photonic-crystal waveguides with high energy and momentum resolution using near-field scanning optical microscopy. 
Intriguingly, we observe additional bands that are not predicted by eigenmode solvers, as was recently demonstrated by Huisman et al. [Phys. Rev. B 86, 155154 (2012)]. 
We study the presence of these additional bands by performing measurements of these bands while varying the incident light power, revealing a non-linear power dependence. 
Here, we demonstrate experimentally and theoretically that the observed additional bands are caused by a waveguide-specific near-field tip effect not previously reported, which can significantly phase-modulate the detected field. 
\end{abstract}
\maketitle

\section{Introduction}

For the investigation of the propagation of light in nanosized stuctures, near-field scanning optical microscopy (NSOM) is a powerful and unique technique as it allows for measurements with a high spatial, energy, and momentum resolution. 
Its ability to tap light out of light-confining structures such as integrated optical waveguides \cite{Balistreri2001, Gersen2005} and cavities, \cite{Mujumdar2007, Lalouat2008}\, makes NSOM an invaluable and popular tool in nanophotonics. \cite{Balistreri2000, Gersen2005}
One very important class of such nanophotonic structures consists of photonic-crystal waveguides, which are two-dimensional photonic-crystal slabs with a line defect wherein the light is guided.
Their importance is found in their unique properties such as their dispersion relation, slow-light propagation, strong confinement of light and thus enhanced light-matter interaction. \cite{ Fan1999,Krauss2007, Joannopoulos2008, LundHansen2008, Topolancik2007, Sapienza2010, Faolain2010}
With NSOM, one has the tool to determine the dispersion relation in these waveguides and thus the bandstructure, spatially map optical pulses, observe slow-light propagation and phenomena such as disorder-induced formation of Anderson localized modes near the band edge, \cite{Engelen2005, Gersen2005a, Gersen2005, Volkov2005, Huisman2012,Huisman2012a}\, which otherwise would not be accessible.

Here, we present our results on measuring the bandstructure of a photonic-crystal waveguide. We show that our measurements reveal additional bands which are not accounted for in eigenmode solvers.
We analyze these new modes by controlling the incident light intensity in power-scaling measurements. 
Their nonlinear scaling behavior is different from the linear response of the known bands.
We explain in detail the origin of these nonlinear bands as a consequence of mode coupling caused by thermal perturbation of the standing waves formed in a finite-size GaAs photonic-crystal waveguide.
We demonstrate that these new bands are in fact not modes belonging to the photonic-crystal waveguide but a measurement artefact caused by the presence of the NSOM tip.

\section{Experimental Details}

\begin{figure}
 \includegraphics[]{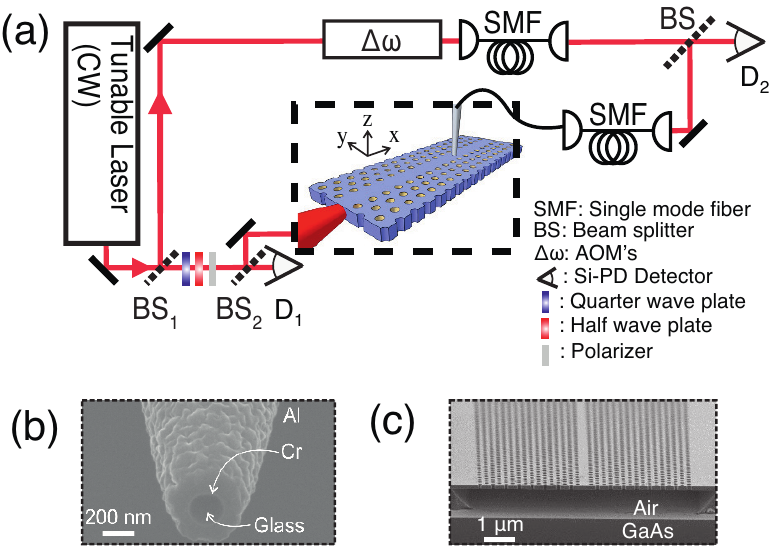}
 \caption{\label{fig:fig1}(Color online) (a) Schematic of the heterodyne near-field scanning optical microscope setup.
 (b) Scanning electron micrograph of the NSOM tip with an aperture of $\approx 160\ \text{nm}$.
 (c) Scanning electron micrography of the GaAs photonic-crystal waveguide membrane structure used in the experiments. The W1 waveguide is formed by the missing row of holes in the center of the photonic-crystal slab.}
\end{figure}

Figure \ref{fig:fig1}(a) shows the experimental setup. 
It is an extended version of the setup described in Ref. [\cite{Huisman2012a}]. 
Therefore, we will concentrate on the implemented improvements. 
A tunable continuous-wave laser (Toptica DLpro 940) with a wavelength range between $907\ \text{and}\ 990\ \text{nm}$ and a linewidth of $100\,\text{kHz}$ is used. In order to extract both amplitude as well as phase out of the NSOM, a heterodyne interferometric detection technique is used. \cite{Balistreri2000}

The signal path is equipped with a motorized combination of $\lambda/4$- and $\lambda/2$-plates, a polarizer, and a photodiode ($\text{D}_1$, see Fig.\,\ref{fig:fig1}(a)). 
The output of $\text{D}_1$ behind the beamsplitter $\text{BS}_2$, \textit{i.e.}, the induced photocurrent, measured as a voltage drop behind a fixed resistor, is measured against a calibrated power meter on the signal input, resulting in a one-to-one mapping of voltage to input power. 
Using this arrangement, we have the ability to tune the input power very accurately over a large range, \textit{i.e.}, $50\,\mu\text{W}$ to $15\,\text{mW}$, throughout the entire wavelength range of the laser.
After this calibration, the signal input light is coupled to the cleaved end facet of a GaAs photonic-crystal waveguide by means of a microscope objective ($\text{NA}=0.55$) and propagates there along the positive $\hat{x}$-direction. 
The polarization of the incident light is kept thereby at approximately $45^{\circ}$ with respect to the normal of the waveguide allowing us to excite both TE- and TM-like modes simultaneously. 
The field pattern is collected $\approx 100\,\mu\text{m}$ away from the coupling facet -- to avoid direct light scattering into the collection part -- using an Al-coated fiber tip with an aperture of $\approx 160\,\text{nm}$ (Fig.\,\ref{fig:fig1}(b) shows a scanning electron micrograph (SEM) of such a tip).
The tip is thereby kept at a fixed distance of $\approx 20\,\text{nm}$ above the sample surface using shear-force feedback control. 
The used height is a trade off between picking up enough evanescent light and disturbing the light field by the tip's presence.
The picked-up light is heterodyned with the local oscillator and detected on a Si photodiode ($\text{D}_2$) allowing to accurately measure amplitude and phase with subwavelength spatial resolution.  

The parameters of the GaAs photonic-crystal slab waveguide, which are also used in the calculations, are extracted from SEM images, such as the one shown in Fig.\,\ref{fig:fig1}(c) and are: pitch size of the triangular lattice $a=240\pm10\,\text{nm}$, normalized hole radius $r/a=0.309\pm0.002\,\text{nm}$, and slab thickness $h=160\pm10\,\text{nm}$. The length of the waveguide $l$ of approximately $1\,\text{mm}$ is derived from optical microscope images. 

\section{Results}  

\begin{figure}
 \includegraphics[]{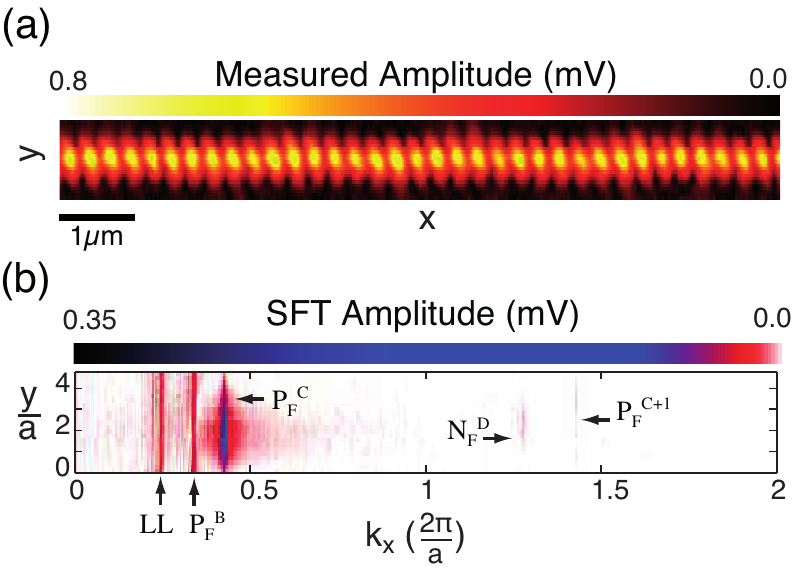}
 \caption{\label{fig:fig2}(Color online) (a) Measured amplitude profile of the photonic-crystal waveguide at an incident wavelength of $\lambda = 964.4 \pm 0.01\,\text{nm}$. 
 (b) Amplitude coefficients of the spatial Fourier transform (SFT) of the measured near-field pattern. Modes marked with 
${\rm P}_{\rm F}^B$, 
${\rm P}_{\rm F}^{\rm C}$,
${\rm N}_{\rm F}^{\rm D}$ and 
${\rm P}_{\rm F}^{\rm C+1}$  correspond to bands responsible for the standing-wave pattern present in Fig.\,2(a).
}
\end{figure}

With our NSOM we map the amplitude of the light field inside the photonic-crystal waveguide with high spatial resolution. 
Figure \ref{fig:fig2}(a) shows a detailed part of such a scan, which extends normally $52\,\mu\text{m} \times 1.2\,\mu\text{m}$ in the $\hat{x}$- and $\hat{y}$-direction, respectively, showing a standing-wave pattern of the light inside the waveguide. 
The wavelength of the light coupled into the waveguide is $\lambda=964.4\pm0.01\,\text{nm}$. 
Analyzing such a spatially resolved measurement further allows the extraction of all the (Bloch-)periodic optical signals in the waveguide by means of a spatial Fourier transform (SFT). \cite{Gersen2005, Huisman2012, Ha2011, Spasenovic2011}
Figure \ref{fig:fig2}(b) shows the result of the SFT performed on our measurement in Fig.\,\ref{fig:fig2}(a), revealing four bands and the light line (LL).

\begin{figure}
\includegraphics[]{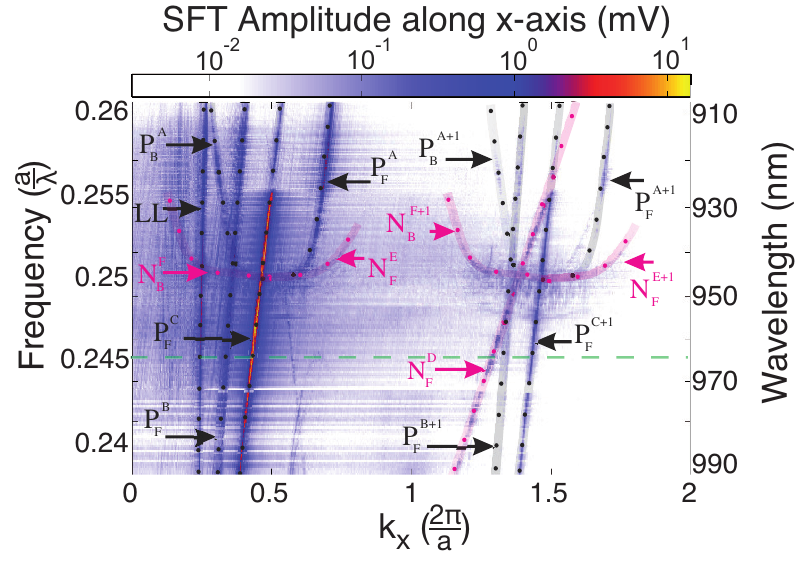}
\caption{\label{fig:fig3}(Color online) Measured bandstructure of a photonic-crystal waveguide. 
The measured bands are labeled from A to F (in the superscript), LL denotes the light line. Furthermore, the results of MPB simulation are inserted as black dots. Modes present in experiment and calculation are labeled P (predicted). Bands measured but not appearing in the calculations are shown in magenta and labeled N (not predicted). 
The subscript denotes backward (B) or forward (F) propagation.
The green dashed line denotes the frequency of more detailed analysis.}
\end{figure}

We can reconstruct the photonic bandstructure of our system very accurately by collecting more near-field images across the whole frequency range of the laser source. Figure \ref{fig:fig3} displays the so reconstructed bandstructure for our photonic-crystal waveguide.
Furthermore, we included the eigenmodes of the system as calculated using the MIT photonic bands eigenmode solver (MPB). \cite{mpb} 
Comparing the results, we can match the measured bands to those in the calculations.
Strikingly, our measurements demonstrate additional bands that are not predicted in the eigenmode solvers.
To be certain that the bands calculated by the MPB eigenmode solver are predicted correctly, we varied several parameters, such as the system size, the unit cell as well as the resolution in the calculations. 
The measured new bands will subsequently be analyzed in more detail.

For clarity, we introduce here the following nomenclature: Measured bands which show also up in the MPB calculations are denoted P (predicted) while the bands measured but not appearing in the calculations are denoted N (not predicted). 
Moreover, from the slope in the bandstructure, as derived from Fig.\,\ref{fig:fig3}, one can derive the group index and energy velocity, thus knowing if it is a forward (F) or backward (B) propagation, which is marked in subscript. 
The band name appears alphabetically in superscript.
For example, a forward propagating predicted mode A is denoted as P$_{\rm F}^{\rm A}$, a backward propagating structure F as N$_{\rm B}^{\rm F}$, and a forward propagating predicted mode C in the second Brillouin zone as P$_{\rm F}^{\rm C+1}$.
Finally the light line appears in its usual abbreviation as LL.
\begin{figure}
\includegraphics[]{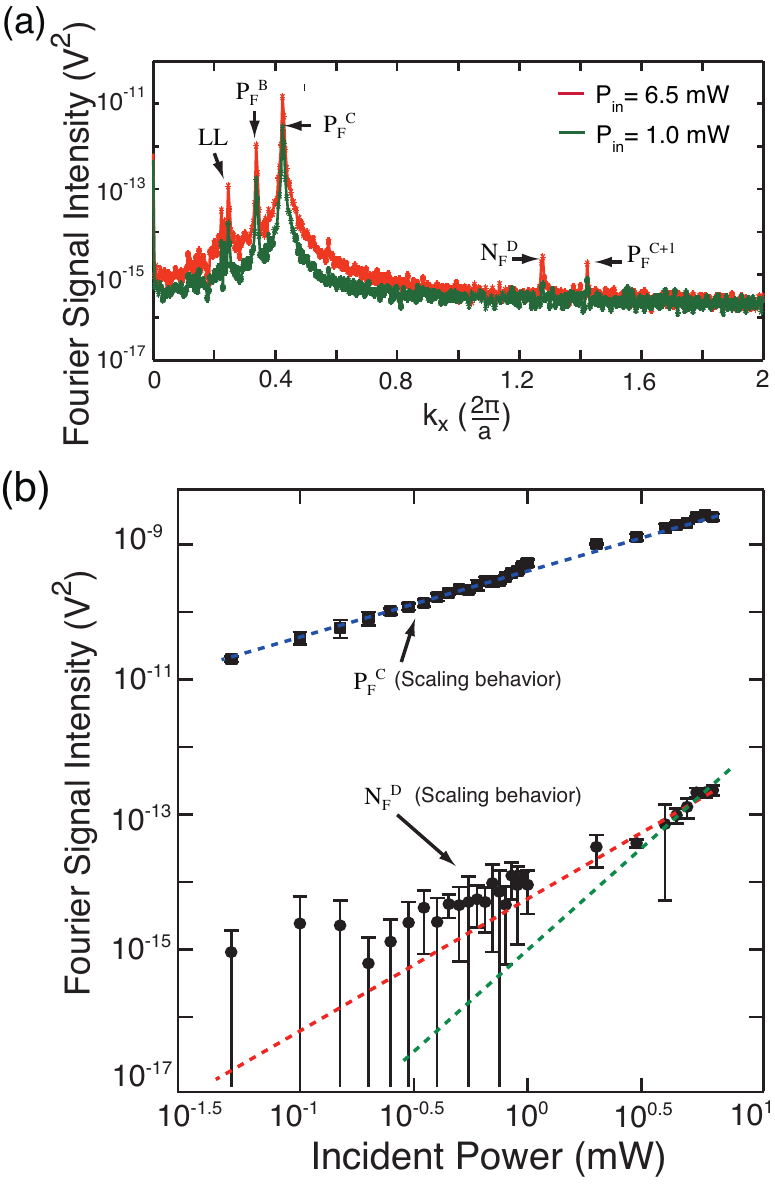}
\caption{\label{fig:fig4}(Color online) Power scaling experiment to measure the intensity-dependent behavior of waveguide bands. 
(a) Total intensity in the Fourier signal (as shown in Fig. \ref{fig:fig2}(b)) as function of the spatial frequency $k_{\text{x}}$. 
The curves represent two different incident power levels: P$_{\text{in}}=6.5\ \text{mW}$: red; P$_{\text{in}}=1.0\ \text{mW}$: green. The bands (LL, P$_{\rm F}^B$, P$_{\rm F}^{\rm C}$, N$_{\rm F}^{\rm D}$ and P$_{\rm F}^{\rm C+1}$) are clearly visible as peaks.
(b) Fourier signal intensity of P$_{\rm F}^{\rm C}$ and N$_{\rm F}^{\rm D}$ measured as a function of incident power. The dotted lines represent linear (blue), quadratic (red), and cubic (green) fits to the data, whereby N$_{\rm F}^{\rm D}$ is fitted only for incident powers exceeding $1\ \text{mW}$ (\textit{cf.} text).
}
\end{figure}  

In Fig.\,\ref{fig:fig3} we identify three new bands, labeled D, E, and F. We can construct these new bands from predicted bands following a general rule: 

\begin{equation}
\label{eqn:eq1}
k_x^{(\text{new\ band)}} = 2 k_x^{(\text{predicted\ band)}} \mp k_x^{(\text{second\ predicted\ band)}}
\end{equation}

\noindent In case of D, E, and F this leads to:
\begin{eqnarray}
\label{eqn:eq2}
\begin{aligned}
k_x({\rm N}_{\rm F}^{\rm D}) &= 2  k_x({\rm P}_{\rm F}^{\rm C}) + k_x({\rm P}_{\rm F}^{\rm C})=3  k_x({\rm P}_{\rm F}^{\rm C})\\
k_x({\rm N}_{\rm F}^{\rm E}) &= 2 k_x({\rm P}_{\rm F}^{\rm A}) - k_x({\rm P}_{\rm F}^{\rm C})\\
k_x({\rm N}_{\rm B}^{\rm F}) &= 2 k_x({\rm P}_{\rm B}^{\rm A}) - k_x({\rm P}_{\rm F}^{\rm C}),\\
\end{aligned}
\end{eqnarray}

\noindent where they are constructed from the predicted modes A and C.

Due to the complexity of the bandstructure, we restrict our further analysis without loss of generality to one specific wavelength below the band edge, \textit{i.e.}, $\lambda=964.4\ \text{nm}$ (green dashed line in Fig.\,\ref{fig:fig3}) and therefore to the new band N$_{\rm F}^{\rm D}$. 
We performed power-dependent measurements to study the origin of these new bands. 
In that respect we took NSOM images as shown in Fig.\,\ref{fig:fig2}(a) in a series of power levels ranging from $0.5-6.5\,\text{mW}$, repeating each measurement 5 times for better statistics. 
From each NSOM image we calculate the SFT (Fig.\,\ref{fig:fig2}(b)) and sum up the signal along the $\hat{y}$-direction.
Figure  \ref{fig:fig4}(a) shows the resulting Fourier signal intensity vs. $k_x$ for two input powers of P$_{\text{in}}=6.5\,\text{mW}$ (red) and P$_{\text{in}}=1.0\,\text{mW}$ (green), respectively. 
The periodic light patterns picked-up by NSOM are thereby clearly visible as peaks above the background signal. 
Furthermore, we can see that the overall intensity of the spectrum changes as a result of the change in power.

To analyze these signals further we plot the intensity of the k component (after background subtraction) as a function of the incident power as is shown exemplarily for the P$_{\rm F}^{\rm C}$ and N$_{\rm F}^{\rm D}$ in Fig.\,\ref{fig:fig4}(b).
Surprisingly, the two bands show a different scaling behavior with the incident power. 
For the predicted mode P$_{\rm F}^{\rm C}$ a linear dependence of the intensity is observed as expected. 
The new band N$_{\rm F}^{\rm D}$ on the other hand shows a highly nonlinear behavior. 
Here, the power law appears to be cubic at powers $\text{P}>3\,\text{mW}$, and quadratic at intermediate powers.
In the following we explain the observed nonlinear behavior of these new bands appearing in the measurement.

\section{Theoretical Model} 
To understand the experimental findings and why the eigenmode calculations cannot predict the measured bands, we first look at the difference between calculations and experiment.
In contrast to the MPB calculations, which assume an infinite sample length and linear material response, our experiment consist of a finite photonic crystal waveguide. Moreover, GaAs is known for its highly nonlinear material response. \cite{Blakemore1987}
Therefore, the eigenmodes of the waveguide as calculated by MPB can only be a starting point for the explanation of the real situation, since it cannot account for any coupling between modes. 

In the following we will show that this measurement-induced coupling mechanism leads to virtual mode coupling and is essential for the observation of new bands in our experiment. 
We restrict our analysis, without loss of generality, thereby to the effects caused by the brightest eigenmode C. 
The same effects will also be present in other (weaker) eigenmodes.
Furthermore, we can omit the Bloch periodicity in the analysis of the electric field because it does not affect the appearance of the observed new bands.

\begin{figure}
 \includegraphics[]{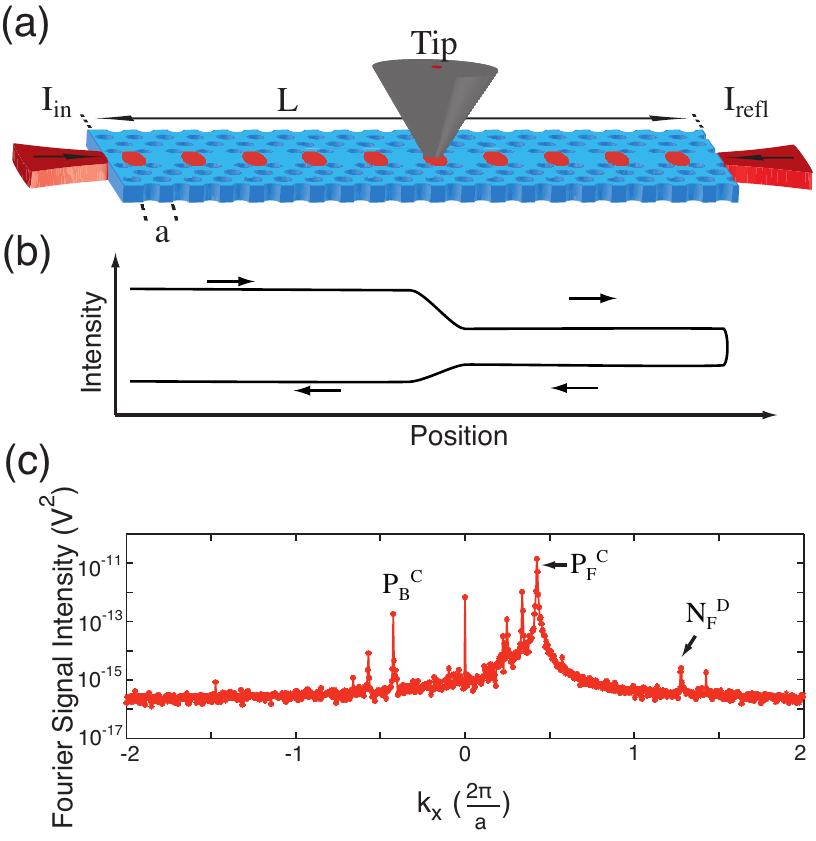}
 \caption{\label{fig:fig5}(Color online) (a) Schematic representation of a standing-wave intensity pattern formed by counter-propagating waves in a finite photonic-crystal waveguide. The signal is then picked-up with the NSOM tip.
(b) Schematic representation of the position-dependent total intensity inside the photonic-crystal waveguide. 
The intensity drop during forward and backward propagation is caused by light scattering and absorption of light due to the presence of the tip.
(c) Intensity of the measured Fourier signal (\textit{cf.} Fig.\,\ref{fig:fig4}(b)), as a function of spatial frequency $k_x$ at an incident power of P$_{\text{in}}=6.5\,\text{mW}$. Modes labeled P$_{\rm B}^{\rm C}$ and P$_{\rm F}^{\rm C}$ correspond to two counter-propagating waves and N$_{\rm F}^{\rm D}$ is the new band appearing in this experiment.}
\end{figure}

In the finite structure, reflections from the end facets of the waveguide will lead to a formation of a standing-wave pattern due to forward ($+k_x$) and backward ($-k_x$) propagating modes, as is schematically depicted in Fig.\,\ref{fig:fig5}(a).
The complete electric field associated with the forward propagating bright eigenmode C at position $r$ and frequency $\omega$ is given by $E_{{\rm P}_{\rm F}^{\rm C}}(r,\omega)= A_{{\rm P}_{\rm F}^{\rm C}} (r,\omega) e^{-ik_c x/a}$, where $A_{{\rm P}_{\rm F}^{\rm C}} (r,\omega)$ is the field amplitude.
The backward propagating mode is then given by $E_{{\rm P}_{\rm B}^{\rm C}}(r,\omega)= A_{{\rm P}_{\rm B}^{\rm C}} (r,\omega) e^{ik_c x/a}$.
The total intensity $I(r,\omega)$ inside the waveguide then becomes

\begin{equation}
\label{eqn:eq3}
I(r,\omega) = I_{\rm BG} + 2| A_{{\rm P}_{\rm B}^{\rm C}} (r,\omega)A_{{\rm P}_{\rm F}^{\rm C}} (r,\omega)|\cos{(2k_c \frac{x}{a}),} 
\end{equation}

\noindent where 
$I_{\text{BG}}$ corresponds to the non-sinusoidal background intensity.
Assuming an intensity dependence for the refractive index and inserting Eq.\,\ref{eqn:eq3} as expression for the intensity, the refractive index $n(r,\omega)= n_{l}(\omega ) + n_{2}(\omega) I(r,\omega) $ can then be written as:

\begin{equation}
\label{eqn:eq4}
\begin{split}
n(r,\omega) &=n_{l}(\omega ) + n_{2}(\omega) I_{\text{BG}} \\
&+n_{2}\cdot2 | A_{{\rm P}_{\rm B}^{\rm C}} (r,\omega)A_{{\rm P}_{\rm F}^{\rm C}} (r,\omega)|\cos{(2k_c \frac{x}{a})},
\end{split}
\end{equation}

\noindent where $n_{l}(\omega)$ is the linear refractive index of GaAs and $n_{2}(\omega)\approx n_{2}$ is the corresponding nonlinear refractive index at the incident optical frequency $\omega$ and intensity $I(r,\omega)$.

We need one more ingredient to explain the appearance of the new bands, namely the perturbative nature of NSOM measurements. 
While scanning across the surface of the waveguide, the tip causes losses predominantly at the maxima of the standing-wave pattern (\textit{cf.} Fig.\,\ref{fig:fig5}(a)).
This, however, is a loss mechanism for the total power inside the waveguide leading to a variation of the refractive index in the waveguide as a function of the position of the tip.
Using $I(r,\omega)$ (Eq.\,\ref{eqn:eq3}) and leaving out the explicit position and frequency dependence of the amplitudes, the perturbation of the refractive index of a large fraction of the waveguide caused by the moving near-field tip can be described through 

\begin{equation}
\begin{split}
\label{eqn:eq6}
& n(r,\omega) = n_{0}[1+ n_{2} I(r,\omega)] \\ 
&= n_{0} [1+ n_{2}I_{\rm BG} + n_{2} \cdot 2| A_{{\rm P}_{\rm B}^{\rm C}} A_{{\rm P}_{\rm F}^{\rm C}}|\cos{(2k_c \frac{x}{a})}]\\
&= n_0+\Delta n.
\end{split}
\end{equation}

\noindent 
Here we assume that the field in the waveguide is dominated by band C as justified by Fig.\,\ref{fig:fig3}.

As a consequence of this refractive-index change an additional phase $\phi$ is introduced to the propagating light. 
The change in phase for any forward propagating mode between the front facet and the NSOM tip is thus

\begin{equation}
\label{eqn:eq7}
\begin{split}
\Delta \phi(r,\omega) &= k_c r \Delta n \\
&= k_c r n_{0} n_{2} [I_{\rm BG} + 2 | A_{{\rm P}_{\rm B}^{\rm C}} A_{{\rm P}_{\rm F}^{\rm C}}|\cos{(2k_c \frac{x}{a})}]\\
&= \phi_{\text{BG}} + k_c r n_{0} n_{2} 2| A_{{\rm P}_{\rm B}^{\rm C}}A_{{\rm P}_{\rm F}^{\rm C}}|\cos{(2k_c \frac{x}{a})},
\end{split}
\end{equation}

\noindent where $\phi_{\rm BG}$ corresponds to the phase term caused by the average temperature rise and hence does not contribute to the new bands.
Backward propagating modes experience a similar phase shift.
The measured field can therefore be described as the unperturbed signal with an additional phase. 
Expanding the phase by a Taylor series results in the general expression for the field detected by the NSOM measurements:
\begin{equation} 
\label{eqn:eq8}
\begin{split}
&E_{det}(r,\omega) = A e^{-ik\frac{x}{a}} e^{-\bigtriangleup \phi(r,\omega)}\\
&= A e^{-ik_c \frac{x}{a}} e^{-\phi_{\rm BG}}\\
&\cdot [1+ i k_c r n_{0} n_{2}\cdot 2 A_{{\rm P}_{\rm B}^{\rm C}} A_{{\rm P}_{\rm F}^{\rm C}} \cos{(2k_c \frac{x}{a})}\\
&+\frac{(i k_c r n_{0} n_{2} \cdot 2 A_{{\rm P}_{\rm B}^{\rm C}} A_{{\rm P}_{\rm F}^{\rm C}} \cos{(2 k_c \frac{x}{a})})^2}{2!}+...]\\
&= \sum_{N=0}^{\infty }A e^{-(ik_c \frac{x}{a}+ \phi_{\rm BG})}\cdot \frac{(e^{-i2k_c \frac{x}{a}}+e^{i2k_c \frac{x}{a}})^N}{2^N N!} \\ 
&\cdot[i k_c r n_{0} n_{2}\cdot 2 A_{{\rm P}_{\rm B}^{\rm C}} A_{{\rm P}_{\rm F}^{\rm C}}]^{N},
\end{split}
\end{equation}  

\noindent where $A$ is the amplitude of any real mode in the waveguide with wave vector $k$. In particular, $A$ could be $A_{{\rm P}_C^{\rm F}}$. Equation \ref{eqn:eq8} gives the general expression of the field associated with mode C of the intensity-perturbed photonic-crystal waveguide. 
Its Fourier transform will reveal all spatial frequencies present in the measured field data. 
The bands will thereby follow the general expression: $k_{det} = 2Nk_c+k$, where $N=0, 1, 2,...$ . 
Pure mode C corresponds to the term $N=0$ in Eq. \ref{eqn:eq8} and the field associated with this mode is $A_{{\rm P}_{\rm F}^{\rm C}}\exp[-ik_c \frac{x}{a}+\phi_{\rm BG}]$.
The intensity is thus given by $|A_{{\rm P}_{\rm F}^{\rm C}}|^{2}$, scaling linearly with the incident intensity as expected for a pure mode.    

A new band as observed in the experiment corresponds to $N=1$ in Eq. \ref{eqn:eq8} and its field is given by

\begin{equation} 
\label{eqn:eq9}
\begin{split}
&E_{N=1}(r,\omega)= i k_c r n_{0} n_{2} \cdot 2 A_{{\rm P}_{\rm B}^{\rm C}} A_{{\rm P}_{\rm F}^{\rm C}} \\ 
&\cdot A_{{\rm P}_{\rm F}^{\rm C}} e^{-(ik_c \frac{x}{a}+\phi_{\rm BG})} \frac{e^{-i2k_c \frac{x}{a}}+e^{i2k_c \frac{x}{a}}}{2}\\
&= A_{{\rm P}_{\rm B}^{\rm C}} A_{{\rm P}_{\rm F}^{\rm C}} A_{{\rm P}_{\rm F}^{\rm C}} e^{\phi_{\rm BG}} i k_c r n_{0} n_{2} \\ 
&\cdot [ e^{(-i3k_c \frac{x}{a})} + e^{(+ik_c \frac{x}{a})}] .
\end{split}
\end{equation}  

It contains bands corresponding to $3k_c$ and $k_c$ and the associated field amplitude for both these bands is proportional to $ A_{{\rm P}_{\rm B}^{\rm C}} \cdot A_{{\rm P}_{\rm F}^{\rm C}} \cdot A_{{\rm P}_{\rm F}^{\rm C}}$. 
In the case of N$_{\rm F}^{\rm D}$ will scale as $|A_{{\rm P}_{\rm B}^{\rm C}}|^2\cdot |A_{{\rm P}_{\rm F}^{\rm C}}|^2\cdot |A_{{\rm P}_{\rm F}^{\rm C}}|^2$, therefore the intensity of the new band should scale cubic with the incident intensity. 
In the same way the new bands E and F can be explained not with mode C but with mode A as perturbed signal field.

\section{Thermal origin of nonlinearity} 
\label{thermalSection}

In principle, several nonlinear effects can be responsible for the refractive index modification as a consequence of an intensity change. The Kerr effect and a heat-induced index change are the most obvious ones.
Our measurements alone do not allow us to dissect the effects, but a thermal origin of the new bands seems the most likely as we will argue below.
With a Kerr coefficient of the order of $10^{-11}\,$cm$^2/$W, \cite{Yuce2012} 1\,mW of power in our waveguide leads to an index of refraction change of less than $10^{-7}$. On a length of a mm this causes a $10^{-4}\,$ rad phase shift.

To understand the effect of heating, let us recall the processes leading to a temperature increase inside the waveguide. 
The photon energy coupled into the waveguide is between $1.25\ \text{and}\ 1.36\ \text{eV}$ and therefore no direct one-photon absorption of the light is possible (band gap of GaAs: $E_{\rm g, GaAs}=1.42\ \text{eV}$).
However, due to the confinement of the light inside the waveguide, the intensity can increase such that the probability of two-photon absorption processes becomes non-negligible. 
Accompanying the photon absorption is a heating up of the sample. 
In addition, contaminations or defects at the slab's surface can lead to additional heating. In fact, we have observed similar waveguides to be destroyed by a runaway thermal degradation at slightly higher excitation powers.
Taking the thermal diffusivity and conductivity of GaAs into account, \cite{Blakemore1987} the time it takes the material to thermalize on the length scale of the wavelength $\lambda$ is of the order of $10^{-8}\ \text{s}$, which means that the induced local heating pattern (due to the standing wave) is completely washed out and therefore homogeneous. As explained before, however, the total intensity depends on the position of the NSOM tip relative to the intensity maxima in the standing light wave. 
The alteration of the band intensity correlates to a temperature change as $\partial_{x} T (r, \omega) = \alpha\cdot \partial_{x} I(r, \omega)$, with $T (r,\omega)$ the absolute temperature of the photonic-crystal waveguide as a function of spatial position $r$ of the near-field tip. 
The resulting refractive index due to the heating is given by $n_{\text{th}} = n_{0}(1+n_{1} T)$,
\noindent where $n_{0}$ and $n_{1}$ for GaAs are 3.255 and $n_{1}=4.5\times 10^{-5}\ {K^{-1}}$, respectively. \cite{Blakemore1987} 
A rough estimate of the temperature rise given the intrinsic absorption coefficient of GaAs and the heat conductivity of the GaAs slab yields a temperature rise of about 1\,K per mW pump power. Given $n_1$, this is three orders of magnitude more than the index change caused by the Kerr effect.

\section{Conclusions and Discussion} 
Our model based on the perturbation caused by a near-field tip correctly explains the appearance of the observed new bands. Hence, these bands are  not optical modes of the waveguide but a waveguide-specific NSOM tip effect not previously observed, where the NSOM tip can cause a large phase shift.
There are several circumstances enabling us to observe the new bands in our experiment: 
1) Use of a direct-band gap material with a large thermal coefficient, 
2) A high thermal isolation provided by the free-standing perforated waveguide samples, and 
3) Phase-sensitive measurement allowing to map and separate individual spatial signals in Fourier space. 
The observed scaling factor (\textit{cf.} Fig.\,\ref{fig:fig4}(b)) matches the one predicted by our model for higher powers.
Competing extrinsic as well as intrinsic refractive index perturbing processes cannot be ruled out, though.
Therefore, even though we successfully explained the observed modal appearance as near-field tip-induced thermal perturbation, we cannot exclude the influence of intensity-dependent nonlinear optical effects, since there has already been experimental evidence of a very large nonlinear optical response measured in InGaAs photonic-crystal waveguide structures, \cite{Cestier2010} yet without clarity on the origin of the effect.

\section{Summary}  
In conclusion, we have observed and explained the existence of new bands in NSOM measurements of photonic crystal waveguides not predicted by standard eigenmode calculations. 
Our experimental results demonstrate an intriguing effect caused by coupling of modes in a finite-sized photonic-crystal waveguide and could be explained as a result of position-dependent tip losses and accompanying temperature changes.

\section{Acknowledgments}  
We thank D.J. Dikken, L. Kuipers, A. Lagendijk,  P. Lodahl, H.L. Offerhaus, S. Stobbe, and W.L. Vos for stimulating discussions and F.B. Segerink and C.A.M. Harteveld for technical support. This work was supported by NWO-nano and FOM, a subsidiary of NWO.

\end{document}